\documentclass[aps,prl,twocolumn,footinbib,reprint]{revtex4}
\usepackage[T1]{fontenc}
\usepackage[small,bf]{caption} 
\usepackage[latin9]{inputenc}
\usepackage{graphicx} 
\usepackage{amssymb}
\usepackage{amsmath}

\usepackage{color}

\begin{document} 
\title{Coexistence of Scale-Invariant States in Incompressible Elastomers}
\author{Evan Hohlfeld} 
\email{evanhohlfeld@gmail.com}
\affiliation{Department of Physics, University of Massachusetts, Amherst, MA 01003, USA} 

\begin{abstract}
Cavitation and sulcification of soft elastomers are two examples of thresholdless, nonlinear instabilities that evade detection by linearization. I show that the onset of such instabilities can be understood as a kind of phase coexistence between multiple scale-invariant states, and I constructively enumerate the possible scale-invariant states of incompressible rubber in two dimensions.  Whereas true phases (like the affine deformations of rubber) are homogeneous, the alternatives are inhomogeneous. In terms of the thermodynamics of solids, both classes of states must generally be given equal consideration.
\end{abstract}

\maketitle

First order phase transitions such as the liquefaction of argon, nematic ordering of the liquid crystal 5CB, and allotropy in iron 
can be understood at a macroscopic scale as coexistences between different scale-free, \emph{homogenous} states.
Recently, soft solids such as a hydrogels, elastomers, and  tissues, were discovered to possess a novel kind of scale-free  instability with an uncanny, though imperfect resemblance to a first order phase transition \cite{Hohlfeld-Mahadevan,Hohlfeld-Mahadevan2012,Chen}.  When a soft solid surface is sufficiently compressed, infinitesimal, sharply creased folds can nucleate and grow in this surface \cite{Hohlfeld-Mahadevan,Chen}. The critical compressive strain for nucleation is independent of the sample shape and marks the coexistence of localized folds, called \emph{sulci}, and a smooth free surface. Further quasistatic deformation causes the nucleated sulci to grow (or shrink) while a quantity of work---which can be interpreted as an energy of transformation---flows into (or out of) each folded region \cite{Hohlfeld-Mahadevan2012}. During this processes, the strain close to each sulcus remains at the coexistence value. Other parallels with phase transitions include the existence of an upper critical strain, or ``surface spinodal,''\cite{Biot,Onuki89,Hohlfeld-Mahadevan2012} as well as metastability in a range of  compression  in which the elastomer is linearly stable, yet has no energy barrier for nonlinear instability \cite{Hohlfeld-Mahadevan}. 

Recent interest in sulcification has focused on the effects of swelling \cite{Tanaka,Hayward,Hayward2010,Zalachas,Arifuzzaman}, growth \cite{Benamar, Suo,Bayly}, applied fields \cite{Wang}, mechanical confinement \cite{Gent,Ghatak,P.-M.-Reis:2009fk,Hohlfeld-Mahadevan,Mora,Suo1}, and  imperfections  \cite{Hutchinson}, on how patterns form \cite{Hohlfeld-Mahadevan2012, Chen, Tallinen}, and on connections to other phenomena such as plastic folding  \cite{Sundaram}. However, the  underlying  instability is only partially understood. In \cite{Thesis} it was suggested that that an isolated sulcus in a critically compressed rubber half-space is a local minimum of energy up to translations and changes of scale. 
Because scaling does not change the magnitude of the deformation gradient, the instability toward forming an infinitesimal sulcus is undetectable by linearization. 
This idea was elaborated on in \cite{Hohlfeld-Mahadevan,Suo1,Hohlfeld-Mahadevan2012}; however, the mechanism of strain localization was not explained. 

A natural, related question is  if there are other instabilities that mimic  or generalize the phase-transition-like aspects of sulcification? One immediate example is cavitation in rubbery solids. Like sulcification, cavitation is undetectable in linearized analysis, and  has a characteristic critical stress  for the nucleation and growth of voids \cite{Ball}. 
Actually, concerns about the existence of undetectable instabilities like sulcification are quite old. 
Historically, Weierstrass was the first to suggest that such instabilities might be  generic features of  energies with derivative nonlinearities \cite{Giaquinta}. Following Weierstrass, abstract necessary and sufficient criterion for stability when accounting for such hidden instabilities have been determined. For  vector fields in one dimension and for scalar fields in any dimension, these criteria are related to certain simple convexity properties of the energy. However, the criteria for vector fields in higher dimensions are \emph{not} simple \cite{Grabovsky}, and impractical if the goal is to  detect hidden instabilities.

\begin{figure}
\includegraphics[width=\columnwidth]{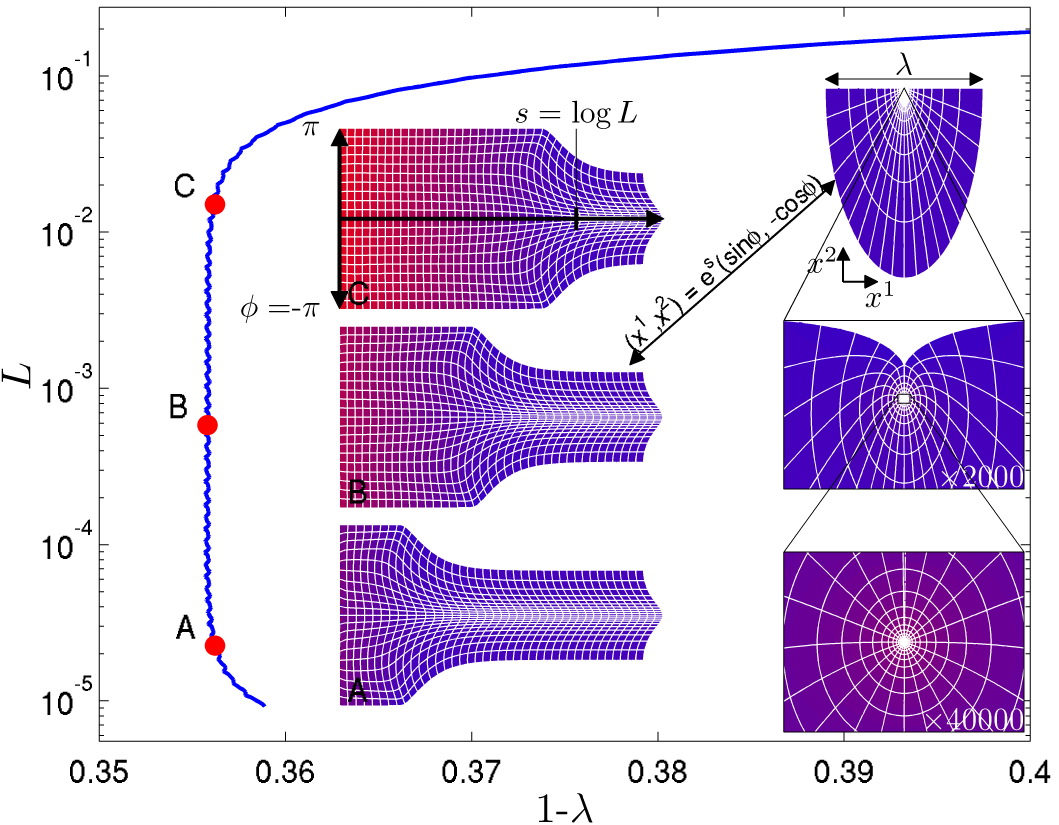}
\caption{\label{fig:fig1}
A sulcus in a unit diameter half-disc compressed to a width $\lambda$ (blowups on right) is mapped by a conformal change of variable  [double headed arrow, also Eqs. (\ref{eq:COV})] to a domain wall in a strip (representative configurations $A-C$)  separating a crease ($s\to -\infty$) and an affine deformation ($s\to +\infty$). Color (blue to red) indicates a linear blow-up of pressure as $s\to -\infty$. The size of the sulcus $L$ is dual to the position of the domain wall on the $s$-axis in the strip. The coexistence compression is $1-\lambda^*\approx0.35.$ The bifurcation diagram (blue line) relates $L^2\propto \lambda-\lambda^*$ \cite{Hohlfeld-Mahadevan2012}; here we view this relationship  as a finite size effect.}
\end{figure}

In this Letter I propose a simple explanation for the similarities between sulcification, cavitation, and  phase transitions, as well as a more practical solution to Weierstrass's stability problem. The idea is that instabilities in scale-invariant systems should result in transitions between scale-invariant states. Then, just as for a phase transition, the coexistence of two such states  determines the threshold of stability. Importantly, these states can be \emph{inhomogeneous}. 
For example, I show that the creased core of a sulcus---like the affine deformation from which it emerges---is scale invariant (in a specific sense), and that the onset of sulcification is given by coexistence of these two scale-invariant states.
This idea is illustrated in Fig. \ref{fig:fig1}, and will be made precise in the course of this Letter.
  I go on to constructively enumerate the scale-invariant states of a model rubbery solid in two dimensions.
These are affine deformations, the cavity, the crease (which exists at a free surface), and a kind of pinched state (which exists at elastomer-elastomer interfaces). 

To frame the discussion, let us consider an infinite sample of incompressible neo-Hookean elastomer. This is a model material for many rubbery solids which are dramatically softer in simple shear than in volumetric compression. It is compactly specified by the Lagrangian density 
\begin{equation}
\mathcal{L}\left(\mathbf{A}\right)=\frac{\mu}{2} A^i_\alpha A_i^\alpha - p\left(\det [A^i_\alpha] -1 \right) \label{eq:NHM}
\end{equation}
where the matrix $\mathbf{A}=\partial \mathbf{x}/\partial \mathbf{X}$ is the gradient of the deformation $\mathbf{x}(\mathbf{X})$ of a reference volume. The first term in  (\ref{eq:NHM}) governs the entropic elasticity of the network chains with shear modulus $\mu$, and the second term, involving the Lagrange multiplier $p$, enforces the nonlinear incompressibility constraint $\det \partial \mathbf{x}/\partial \mathbf{X}=1$. 
The Lagrangian density $\mathcal{L}$ is invariant under separate rigid body motions of $\mathbf{x}$ and $\mathbf{X}$. 
Because $\mathcal{L}$ is a function of the dimensionless deformation gradient $\mathbf{A}$, it is also invariant under the scale transformation $\{\mathbf{x},\mathbf{X}\}\to \{L\mathbf{x},L\mathbf{X}\}$ for any scale factor $L>0$. 

As a preliminary exercise, it is easy to check that any homogenous deformation gradient $\mathbf{A}$ defines a scale invariant state, which in this case is a phase. Then, if 
 the free energy density $\mathcal{F}(\mathbf{A})$ had multiple local minima, 
 one could consider the coexistence of any two phases.
 The compatibility constraint that $\mathbf{A}=\partial \mathbf{x}/\partial \mathbf{X}$ tightly restricts which phases can coexist. 
 Because of  this constraint, phase boundaries 
 can involve at most a rank-one jump in $\mathbf{A}$. That is if $\mathbf{A}_1$ and $\mathbf{A}_2$ are the two phases then $\mathbf{A}_1-\mathbf{A}_2=\mathbf{a}\otimes \mathbf{N}$ where $\mathbf{N}$ is the interface normal and $\mathbf{a}$ is some vector. (Otherwise the elastomer  would tear along the phase boundary.) 
 For any two compatible phases one can compute that $\det\mathbf{A}_1= \det \mathbf{A}_2$ and that $\mathcal{F}$ is a convex function along the chord joining $\mathbf{A}_1$ to $\mathbf{A}_2$ (i.e. $\mathcal{F}$ is rank-one convex). Recalling Maxwell's construction for the van der Waals gas \cite{Pathria}, we infer that  there can be no energy barrier between \emph{compatible} phases of the elastomer, and thus no coexistence of distinct, homogenous states.  

Because of incompressibility, $\mathcal{F}$ is not fully convex \footnote{Consider rotating a homogeneous volume by $\pi$ so that $\mathbf{A}\to-\mathbf{A}$; the chord joining these states passes though $\mathbf{A}=\mathbf{0}$, which is a point of infinite energy.}.
Two manifestations of this nonconvexity are Biot's scale invariant linear instabilities at free surfaces and elastomer-elastomer interfaces \cite{Biot}. At critical compressions, the speeds of Rayleigh or Stoneley waves (respectively) vanish. A vanishing wave speed typically indicates a phase transition (as it does in fluids and  jammed granular systems \cite{Isostatic}),  but as  Biot's instabilities cannot result in phase transitions, the alternative states to the unstable affine deformation must be scale invariant and \emph{inhomogeneous}. 
Our objective now is to enumerate all possible scale invariant states of model \eqref{eq:NHM}, not just homogenous phases. We focus on two dimensions, i.e. plane strain.

When seeking  solutions that exhibit a certain symmetry, it is  helpful to choose a coordinate system that  possesses the same symmetry. For scale symmetry the appropriate coordinates are logarithmic polar coordinates. These are a conformal transformation from the standard  material and lab Cartesian coordinate systems $(X^1,\,X^2)$ and  $(x^1,\, x^2)$, respectively. 
We define the new coordinates systems $(T,\Theta)$ and $(s,\phi)$ by the relations 
\begin{subequations}
\begin{align}
X^1&=\cos(\Theta)e^T, &\quad X^2&=\sin(\Theta)e^T,\\
x^1&=\cos(\phi)e^s, &\quad x^2&=\sin(\phi)e^s.
\end{align}
\label{eq:COV}
\end{subequations}
(I.e. $T=\log R,$ the radial coordinate in the reference body, and likewise for $s$. See insets in Fig. \ref{fig:fig1}.)
This  transformation maps an annulus in the the Cartesian coordinates to a finite strip in the logarithmic-polar coordinates as illustrated in Fig. \ref{fig:fig1}; an infinite system is mapped to an infinite strip. 
More importantly, this coordinate transformation maps rescaling in the Cartesian coordinates to simultaneous translation in $s$ and $T$.

In the new coordinates (and units where $\mu=1$) 
the Euler-Lagrange equations are
\begin{subequations}
\begin{align}
-&\left(\frac{\partial^2 s}{\partial T^2} + \frac{\partial^2 s}{\partial \Theta^2}\right) + \left(\frac{\partial \phi}{\partial T}\right)^2 + \left(\frac{\partial \phi}{\partial \Theta}\right)^2 - \left(\frac{\partial s}{\partial T}\right)^2\notag\\
 &- \left(\frac{\partial s}{\partial \Theta}\right)^2 - \frac{\partial p}{\partial T}\frac{\partial \phi}{\partial \Theta} + \frac{\partial p}{\partial \Theta}\frac{\partial \phi}{\partial T}=0\label{eq:xformed-1a}\\
 &\notag\\
-&\left(\frac{\partial^2 \phi}{\partial T^2} + \frac{\partial^2 \phi}{\partial \Theta^2}\right) - 2 \left(\frac{\partial \phi}{\partial T}\frac{\partial s}{\partial T} + \frac{\partial \phi}{\partial \Theta}\frac{\partial s}{\partial \Theta}\right)\notag\\ &+  \frac{\partial p}{\partial T}\frac{\partial s}{\partial \Theta} - \frac{\partial p}{\partial \Theta}\frac{\partial s}{\partial T}=0,\label{eq:xformed-1b}
\end{align}
\label{eq:xformed-1}
\end{subequations}
and the  incompressibility constraint becomes
\addtocounter{equation}{-1}
\begin{subequations}
\setcounter{equation}{2}
\begin{equation}
\frac{\partial s}{\partial T}\frac{\partial \phi}{\partial \Theta} - \frac{\partial s}{\partial \Theta}\frac{\partial \phi}{\partial T} = e^{2(T-s)}.\label{eq:xformed-2}
\end{equation}
\label{eq:xformed}
\end{subequations}
The second order terms in Eqs. \eqref{eq:xformed} as well as those involving the pressure are the same as would appear in a Cartesian system. The  terms which are quadratic in  $s$ and $\phi$ arise from the Christoffel symbol for the coordinate transformation. Because of scale symmetry, the conformal factor for this transformation cancels, and one is left with a system which is invariant under simultaneous translations of $s$ and $T$, i.e. rescaling. 
Effects that introduce a scale, e.g.  strain gradient terms or external potentials, would be multiplied by factors  of $e^{\pm T}$ when written in the new coordinates, and so have negligible influence  at intermediate values of $T$ as we take the length of the strip to infinity. 

Scale symmetric solutions to Eqs. \eqref{eq:xformed} must be $(T,s)$-translation invariant.
Perfect symmetry turns out to be excessively stringent, and so let us weaken this notion to asymptotic translation invariance as either $T\to \infty$ or as $T\to -\infty$. More precisely, we seek asymptotic solutions to Eqs. \eqref{eq:xformed} for which the second derivatives in $T$  approach zero as $\pm T\to \infty$ for each $\Theta$ and are negligible compared to the remaining terms, uniformly in $\Theta$. In this way, scale symmetry reduces the order of Eqs. \eqref{eq:xformed}.

\paragraph{Affine deformations.} Based on the assumption of vanishing second derivatives, we can infer the asymptotic scaling $s\sim aT$, $\phi \sim bT$ for constants $a$ and $b$. Configurations with $b\ne 0$ involve an infinite degree of winding of the elastomer about the origin, and so are inaccessible via a small displacement.  Considering the incompressibility constraint [Eq. \eqref{eq:xformed-2}] when $T\to \infty$, we find that $a=1$; moreover, the only  solutions  are affine deformations. Up to separate rotations in the material and lab frames, these are parametrized by the principle stretch $\lambda$ and are given by   
\begin{subequations}
\begin{align}
s&=T + \frac{1}{2}\log\left(\frac{1}{\lambda^2}\sin^2\Theta + \lambda^2\cos^2\Theta\right),\\ \phi &= \arctan\left[\frac{1}{\lambda^2}\tan \Theta\right].
\end{align}
\label{eq:affine}
\end{subequations}

The form of Eqs. \eqref{eq:affine} tightly restricts what kinds of scale invariant states can coexist. For example the solution at an interface, e.g., at a corner subtending an angle $\alpha$, must have the form in Eqs. \eqref{eq:affine} on both sides of the dividing surface. Seeking a solution which has the bilateral symmetry of the interface fixes the relative rotations of each side. This leaves as unknowns the two stretch parameters and a constant pressure jump across the interface. The solution, however, must satisfy four continuity conditions at the interface (for two components of displacement and two components of stress); hence, scale-invariant interface solutions can only exist for special angles $\alpha,$ in particular for a flat interface, $\alpha=\pi$.


\paragraph{Cavity.} We find additional solutions to Eqs. \eqref{eq:xformed} when $T\to-\infty$. From Eq. \eqref{eq:xformed-2} we can now have either $a=1$ or $a=0$. The latter situation corresponds to cavitation.  Taking the origin to be a generic interior point, an exact solution exhibiting the coexistence of a  cavity with an affine deformation is $s=\frac{1}{2}\log(e^{2T} +c )$, $\phi=\Theta$, and $p=T - \frac{1}{2}\log(e^{2T} + c) + \frac{c}{2}(e^{2T} + c)^{-1}$. For large $-T$ this solution asymptotes to $a=0$ because $s \to \frac{1}{2}\log(c)$ (i.e. the radius of the cavity is $\sqrt{c}$). For large $T$, $s\sim T - \frac{c}{2}e^{-2T}$, i.e. $a=1$. The crossover region resembles an exponentially localized domain wall with a movable location controlled by the constant $c$.

It is easy to check the cavity solution satisfies our notion of asymptotic scale invariance since second $T$-derivatives are indeed negligible for large $|T|$. The cavity solution also exhibits an important property shared by all coexistence solutions, which is that  joint translation of $s$ and $T$ produces another inequivalent, but energetically degenerate solution \footnote{This phenomenon could be called spontaneous breaking of scale symmetry.}. As the domain wall traverses the strip from left to right (i.e. as $c$ increases or $s$ and $T$ are shifted), the corresponding cavity grows from infinitesimal to infinite size.

The $T\to\infty$ asymptote of the cavity exhibits another generic property of coexistence solutions. In ordinary radial coordinates we  would find $r\sim R + c/2R$.  The algebraic tail of the displacement $r-R$ decays slowly enough to extract energy from any prestress in the far field, resulting in a finite energy of transformation (see \cite{Hohlfeld-Mahadevan2012}).

Fixing the hydrostatic pressure in the cavity, i.e. as $T\to -\infty$, also fixes the pressure as $T\to\infty,$ which determines the coexistence conditions. While this two dimensional cavity solution has infinite energy and coexists with affine deformation at  at infinite hydrostatic tension, higher dimensional cavities have finite energy and a finite coexistence pressure \cite{Ball}. 

\paragraph{Crease.} We can find a third distinct solution to Eqs. \eqref{eq:xformed} by fixing the origin to point on a free surface; in the transformed coordinates the boundary conditions at this surface [which is mapped  to the pair of lines $\Theta=\pm (\pi/2)$] are 
\begin{equation}
\frac{\partial s}{\partial \Theta} - p\frac{\partial \phi}{\partial T}=0,\quad
\frac{\partial \phi}{\partial \Theta} + p\frac{\partial s}{\partial T}=0.\label{eq:BCs}
\end{equation}
At such a point, we set $s=T + g(\Theta)$ and $\partial \phi/\partial T=0$ in Eqs. \eqref{eq:xformed-1} and derive the ordinary differential equation
\begin{equation}
-\frac{\partial^2 g}{\partial \Theta^2} - \left(\frac{\partial g}{\partial \Theta}\right)^2 +e^{-4g}  - \frac{\partial p}{\partial T}e^{-2g} = 1.\label{eq:crease-eq}
\end{equation}
The solutions of this equation with $\partial p/\partial T\ne 0$ describe creases with different opening angles and boundary conditions on the free surface. Because of the diverging pressure the only solution which satisfies the boundary conditions \eqref{eq:BCs} closes to self-contact. This solution is given by the functions $s=T-\frac{1}{2}\log 2$, $\phi=2 \Theta$, $p=\frac{3}{2}T$  \cite{Silling}.   The crease is scale invariant in the same sense as the cavity. Because the contact pressure becomes positive (tensile) at some value of $T$, the crease must eventually convert to an affine deformation for large $T$.

I obtained the domain wall solution that interpolates between the crease and the affine deformations numerically using  finite element and continuation methods as in Ref. \cite{Hohlfeld-Mahadevan}. 
I imposed these two deformations as boundary conditions on opposite ends  of a long strip in the $T-\Theta$ plane and fixed a relative shift in  $x^2$ to eliminate normal forces on these ends. Coexistence between the crease and an affine deformation occurs for the compression $1-\lambda^*\approx 0.35$ (see Fig. \ref{fig:fig1}), which is similar to what has been reported elsewhere \cite{Thesis,Hohlfeld-Mahadevan, Suo1}. Mapping back to Cartesian coordinates, this domain wall becomes the sulcus. 

\paragraph{Pinch.} We obtain the final scale-invariant deformation admitted by model \eqref{eq:NHM} that can coexist with an affine state in two dimensions
by considering an interface between elastomers (denoted by subscripts $+/-$) with shear moduli $\mu<1$ and $1$ (respectively).  Fixing the origin at a point on the interface (here taken to be the lines $\Theta=0,\pi$), the boundary conditions are 
\begin{subequations}
\begin{align}
\mu\frac{\partial s_+}{\partial \Theta} - p_+\frac{\partial \phi_+}{\partial T}&=\frac{\partial s_-}{\partial \Theta} - p_-\frac{\partial \phi_-}{\partial T}\label{eq:shear-bc-0},\\
\mu\frac{\partial \phi_+}{\partial \Theta} + p_+\frac{\partial s_+}{\partial T}&=\frac{\partial \phi_-}{\partial \Theta} + p_-\frac{\partial s_-}{\partial T}\label{eq:normal-force-bc}.
\end{align}
\label{eq:interface-bc-mt}
\end{subequations}
I obtained an asymptotic solution to Eqs. \eqref{eq:xformed} with $a=1$ as $T\to-\infty$  by a solving  an ODE similar to Eq. \eqref{eq:crease-eq} (details are given in the \emph{Supplementary Material}). Now however, the leading $T$-derivatives in Eqs. \eqref{eq:xformed} are secular perturbations to the leading ODE. These perturbations result in the formation of a boundary layer at the material interface and are crucial to the ability to satisfy the boundary conditions in Eqs. \eqref{eq:interface-bc-mt}. The asymptotic form of the  solution is 
\begin{subequations}
\begin{align}
s_\pm&\sim T + \frac{1}{2}\log(A_\pm) + \log\sin\Theta\\
\phi_\pm& \sim A_\pm^{-1}\cot \Theta \pm\frac{\pi}{2}
\end{align}
 for $ T^{-\frac{1}{2}}\ll\pm\Theta\le \pm\frac{\pi}{2}$
 and
 \begin{align}
 s_\pm&\sim T+ \frac{1}{2}\log\left[\frac{1}{A_\pm} + A_\pm\left(\Theta - B_\pm\right)^2\right]\\
 \phi_\pm &\sim \arctan \left[A_\pm\left(\Theta - B_\pm\right)\right]
 \end{align}
 \label{eq:pinch-solution}
 \end{subequations}
  for $0<\pm\Theta\ll1$ (with a symmetric boundary layer at $\Theta=\pi$).
   Here, $A_-=\mu \frac{2}{\pi^2}T^2$ and  $B_-=  \frac{\pi}{2}\left(\frac{1}{\mu} - 1\right)\frac{1}{T}$, and $A_+=\frac{1}{\mu^2}A_-$ and $B_+=\mu B_-$. The isotropic part of the pressure $p_\pm\sim -2\frac{1+\mu}{\pi^2}\frac{T^2}{\log(-T)}$. This solution is asymptotically scale invariant in the same sense as the cavity and crease since the second partial derivatives of $s$ and $\phi$ with respect to $T$ tend to zero for every $\Theta$.  The convergence is not uniform, however, because of the boundary layer at the interface. Within this layer, second $T$ derivatives are $O(1)$ at $|\Theta- B_\pm|\sim A_\pm^{-1}$, but are nevertheless negligible compared to the other terms in Eqs. \eqref{eq:xformed}. Outside the boundary layer, the entire elastomer is pinched to $\phi_\pm=\pm\frac{\pi}{2}$.  Considering the boundary layer, we find that the area fraction of the stiffer material in any disc of radius $e^s$ tends to unity as $s\to -\infty$. 
 Because the elastomer interface can sustain shear, the coexistence surface for the pinch and the affine deformation is more complicated than for the crease (which is specified by a single value of $\lambda$). I will discuss this surface elsewhere. Notice that the pinch \emph{does not} continuously deform to a crease as $\mu\to 0$.

I have argued that the scale-free nonlinear instabilities exhibited by 
elastomers
can be understood in a way analogous to how first order phase transitions are studied. The first step is to identify the accessible scale-invariant sates, and the second step is to find coexistence conditions by obtaining domain-wall-like solutions. As scale symmetry reduces the first step to solving an ODE, we can be confident that the enumeration of states presented here is complete 
for the two dimensional model considered.
 It is natural and straightforward to generalize this approach to higher dimensions in a hierarchical fashion: Localized states in two dimensions become \emph{linelike} states in three dimensions, and new localized states 
 could
 emerge as well.  The coexistence of a three dimensional localized state with an affine deformation is entirely analogous to coexistence in two dimensions. However the coexistence of a line-like state with an affine deformation raises interesting questions. For example, a line-like state should have soft undulation and elongation modes in addition to soft scaling and translation modes. These additional modes might change the exponent governing the bifurcation discussed in \cite{Hohlfeld-Mahadevan2012}.

It is also interesting to reverse the analogy with a phase transition pursued here, and consider the thermodynamic consequences of the existence of multiple, possibly inhomogeneous scale invariant states.  From such a point of view, there is no difference between such a state and an ordinary phase: In an infinite system, the free energy is singular at the coexistence point, and the material begins to transform from one scale invariant state to the other, whether or not it is a proper phase or the ``phase boundary'' has molecular dimensions. Echoing Weierstrass's concerns, it is reasonable to expect that the existence of such alternative states is a generic feature of systems with  derivative nonlinearities, for example solids. Two newly discovered instabilities to which the analysis presented here might be applied are the formation of surface folds in nematic elastomers \cite{Islam}, and an unusual nucleation process observed
 in
 a nominally thermodynamically stable alloy that also seems to involve a ``diffuse phase boundary'' separating incompatible phases \cite{Ball-Koumatos-Seiner}.

I acknowledge support by NSF-MRSEC on Polymers at UMass. I am grateful for  discussions with B. Davidovitch,  J. Hanna, S. Mandre,  and B. Svistunov. I also thank one of the anonymous referees for detailed comments and thought provoking questions.

\bibliographystyle{plain} 

\onecolumngrid
\appendix

\begin{center}
\Large Supplementary Material for ``Coexistence of scale invariant states in incompressible elastomers''
\end{center}

\begin{center}
\large E. Hohlfeld
\end{center}

In this Supplementary Material I derive the asymptotic solution to  Eqs. \eqref{eq:xformed} in the main text given by Eqs. \eqref{eq:pinch-solution} as $T\to -\infty$. I use a slightly different notation than in the main text, here the $+/-$ elastomers are referred to as two  and one respectively. Also the angular variables $\Theta$ and $\phi$  are defined here  so that the interface is mapped to  the lines $\Theta=\pm\frac{\pi}{2}$.

\subsection{Governing ODE in the limit $T\to -\infty$}

First, let us write
\[
 g(\Theta,T) = s-T, 
\] 
and assume that $\partial s/\partial T\to 1$ and $\partial \phi/\partial T \to 0$ as $T\to -\infty$. From the incompressibility constraint, Eq. \eqref{eq:xformed-2}, we identify the dominant balance
\begin{equation}
\frac{\partial \phi}{\partial \Theta} \sim e^{-2g}.\label{eq:phi-eq}
\end{equation}
Substituting Eq. \eqref{eq:phi-eq} into Eq. \eqref{eq:xformed-1a} we derive the equation
\begin{equation}
-\frac{\partial^2 g}{\partial \Theta^2}  + e^{-4g} - 1 - \left(\frac{\partial g}{\partial \Theta}\right)^2 - \frac{\partial p}{\partial T}e^{-2g}=0.\label{eq:g-eq}
\end{equation}
As Eq. \eqref{eq:phi-eq} implies
\[
\frac{\partial^2 \phi}{\partial \Theta^2} + 2\frac{\partial \phi}{\partial \Theta}\frac{\partial g}{\partial \Theta} = 0,
\]
substitution of Eq. \eqref{eq:phi-eq} into Eq. \eqref{eq:xformed-1b} results in the equation for the pressure,
\begin{equation}
\frac{\partial p}{\partial \Theta} = \frac{\partial p}{\partial T}\frac{\partial g}{\partial \Theta}.\label{eq:p-eq}
\end{equation}
Eq. \eqref{eq:p-eq} shows that variation of the pressure with $T$ induces substantial variation of the pressure with $\Theta$. Is also shows that variations in $p(\Theta,T)$ about its $\Theta$-average are relatively small as  $T\to -\infty$. That is, as $T\to -\infty$, we can approximate $p(\Theta,T)$ as a piecewise function which is constant in each material half-plane (or infinite half-strip in the $\Theta$-$T$ coordinate system).

Substitution of Eq. \eqref{eq:p-eq}  into Eq. \eqref{eq:xformed-1a} converts the final term in the latter equation into the expression
\begin{equation}
\frac{\partial p}{\partial T}\frac{\partial g}{\partial \Theta}\frac{\partial \phi}{\partial T}.\label{eq:pressure-term}
\end{equation}
As we do not yet know the relative magnitudes of $\partial p/\partial T$ and $\partial \phi/\partial T$, the term \eqref{eq:pressure-term} is potentially significant compared to the terms retained in Eq. \eqref{eq:g-eq}. We verify that it is harmless by examining the first correction to the leading expression for $\partial \phi /\partial \Theta$ given by Eq. \eqref{eq:phi-eq}. This correction is simply the neglected term in the expression for the incompressibility constraint, Eq.  \eqref{eq:xformed-2}, so that
\begin{equation}
\frac{\partial \phi}{\partial \Theta}\sim e^{-2g} + \frac{\partial g}{\partial \Theta}\frac{\partial \phi}{\partial T}.\label{eq:new-phi-eq}
\end{equation}
When this new expression is substituted into the next-to-last term of Eq. \eqref{eq:xformed-1a} we obtain exact cancelation of the potentially troublesome term given by Eq. \eqref{eq:pressure-term}. The other contributions to Eq. \eqref{eq:g-eq} arising from the additional term  in Eq. \eqref{eq:new-phi-eq} are sub-dominant.  We therefore infer that the solution to Eq. \eqref{eq:g-eq} will yield the correct asymptotic behavior of $g(\Theta,T)$ as $T\to -\infty$.

Our first guess is that $g(\Theta,T)$ approaches a finite limit as $T\to -\infty,$ as it does for the crease.   However, extensive searching (via shooting) for a solution to Eq. \eqref{eq:g-eq} with boundary conditions \eqref{eq:interface-bc-mt} was fruitless,  suggesting that this assumption about $g$ is  flawed, and we must consider the possibility that $g$ diverges (at least somewhere) as $T\to -\infty$. The subsequent computations will reveal that $g(\Theta,T)$ diverges for a.e. $\Theta$, except for a narrow boundary layer containing the interface between the two materials.  

\subsection{Boundary layer analysis of Eq. \eqref{eq:g-eq}}
\subsubsection{Leading order}

Our first step to solving Eq. \eqref{eq:g-eq} is to introduce a change of variable for the dependent variable,
\[
u=e^g,\quad \frac{\partial u}{\partial \Theta} = u\frac{\partial g}{\partial \Theta},\quad \frac{\partial^2g}{\partial \Theta^2} = -\frac{1}{u^2}\left(\frac{\partial u}{\partial \Theta}\right)^2 + \frac{1}{u}\frac{\partial^2 u}{\partial \Theta^2}.
\]
Using this new variable, Eq. \eqref{eq:g-eq} is transformed into
\begin{equation}
\frac{\partial^2 u}{\partial \Theta^2}=u^{-3} - \frac{\partial p}{\partial T}u^{-1} -  u.\label{eq:first-integral-0}
\end{equation}
We can integrate this equation for $u$ once to compute
\begin{equation}
\left(\frac{\partial u}{\partial \Theta}\right)^2 = -\frac{1}{2}u^{-2}  - 2\frac{\partial p}{\partial T}\log(u) - u^2 + A(T),\label{eq:first-integral}
\end{equation}
where $A(T)$ is a new constant of integration. We will solve for $u$ as $A$ becomes asymptotically large.

To organize the calculation let us  define  the rescaled variable
\[
v = \frac{u}{\sqrt{A}}.
\]
In terms of this variable, Eq. \eqref{eq:first-integral} takes the form 
\begin{equation}
\left(\frac{\partial v}{\partial \Theta}\right)^2 = 1 -  v^2 - \frac{1}{A^2 v^2} - 2\frac{1}{A}\frac{\partial p}{\partial T}\log\left(\sqrt{A}v\right).\label{eq:first-integral-outer}
\end{equation}
To leading order in $A^{-1}$, the solution to Eq. \eqref{eq:first-integral-outer} is
\[
v_0 = \cos\left(\Theta +  \Theta_0\right),
\]
where $\Theta_0$ is a constant of integration which is required to be zero by symmetry. Returning to the variables $g$ and $\phi$, we find the leading solution
\begin{align}
g&\sim \log\cos\left(\Theta\right) + \frac{1}{2}\log(A),\\
\phi&\sim \frac{1}{A}\tan \left(\Theta\right)
\end{align}
as $A\to\infty$. 


The  approximation $v\approx v_0$ is of doubtful validity for $|\Theta|\approx \frac{\pi}{2}$ since the notionally small terms in Eq. \eqref{eq:first-integral-outer} (these are the terms which formally diminish in magnitude as $A\to\infty$) diverge as $|\Theta|\to \frac{\pi}{2}$. This divergence suggests that different terms dominate in Eq. \eqref{eq:first-integral} for small $\epsilon_\pm=\Theta \pm \frac{\pi}{2}$. The most singular term is the first term on the right side of Eq. \eqref{eq:first-integral}. Consideration of this term motivates a different scaling of the independent and dependent variables:
\[
w = u\sqrt{A},\quad \xi = A\Theta.
\]
In terms of these variables, Eq. \eqref{eq:first-integral} takes the form
\begin{equation}
\left(\frac{\partial w}{\partial \xi}\right)^2 =1 -w^{-2} - 2\frac{1}{A}\frac{\partial p}{\partial T}\log\left(\frac{w}{\sqrt{A}}\right) - \frac{w^2}{A^2}.\label{eq:first-integral-inner}
\end{equation}
The leading order in $A^{-1}$, the solution to Eq. \eqref{eq:first-integral-inner} is
\[
w=\sqrt{1 + \left(\xi + b\right)^2},
\]
where $b$ is a constant of integration. Transforming $w$ back to $g$ and $\phi$, we find
\begin{align}
g&\sim \frac{1}{2}\log\left[\frac{1}{A} + A\left(\Theta \pm \frac{\pi}{2} - B_\pm\right)\right],\\
\phi& \sim \phi_{\pm} +  \arctan\left[A\left(\Theta - B_\pm\right)\right]
\end{align}
as $A\to\infty$. Here the constant of integration $B_\pm = -\frac{b}{A} \pm \frac{\pi}{2}$. 

We determine the constants $B_\pm$ by the technique of asymptotic matching.
 The perturbation terms in the outer and inner problems specified by Eqs. \eqref{eq:first-integral-outer} and \eqref{eq:first-integral-inner} are simultaneously small in the overlap region defined by
\[
\frac{1}{A} \ll v,\quad w\ll A.
\] 
Substituting the leading solutions in the inner and outer regions we  find that both asymptotes are valid for 
\[
\frac{1}{A}\ll \left|\Theta\pm \frac{\pi}{2}\right| \ll 1.
\]
Expanding the inner solution for large $\xi-b$ and the outer solution for small $\Theta\pm\frac{\pi}{2}$ we find
\[
u\sim \sqrt{A}\left|\Theta\pm \frac{\pi}{2}-B_\pm\right| + O\left(\frac{1}{A^{\frac{3}{2}}|\Theta\pm \frac{\pi}{2}-B_\pm|}\right)
\]
in the inner region and
\[
u\sim \pm\sqrt{A}\left(\Theta \pm \frac{\pi}{2}\right)+ O\left[\sqrt{A}\left(\sqrt{\beta}\Theta \pm \frac{\pi}{2}\right)^3\right]
\]
in the outer region.

Matching reveals that $B_{\pm}\to 0$ 
 as $T\to -\infty$. However the \emph{rate} at which $B_\pm$ converges to zero is important in our ability to satisfy the boundary conditions at the interface. We find this rate by calculating the solutions to the inner and outer problems to   next order $A^{-1}$.

\subsubsection{Next-to-leading order in the outer region}

Corrections to $v_0$ in the outer region can be computed using perturbation theory. One source of perturbation terms is already apparent in Eq. \eqref{eq:first-integral-outer}. Another term arrises from the $T$-derivatives in Eq. \eqref{eq:xformed-1a} by expanding 
\[
\left(1+ \frac{\partial g_0}{\partial T}\right)^2\approx 1 + \frac{1}{A}\frac{\partial A}{\partial T}.
\]
This term modifies Eq. \eqref{eq:g-eq} to
\begin{equation*}
-\frac{\partial^2 g}{\partial \Theta^2} - \left(\frac{\partial g}{\partial \Theta}\right)^2 + e^{-4g} - \frac{\partial p}{\partial T}e^{-2g}-\beta = 0
\end{equation*}
where
\[
\beta = 1 + \frac{1}{A}\frac{\partial A}{\partial T}.
\]
We can account for this perturbation  by a simple rescaling of the independent variable in our leading order solution $v_0$. In particular if we replace
\[
v_0=\cos(\Theta)\to \cos\left(\sqrt{\beta}\Theta\right),
\]
then all that remains is to compute the effects of the subdominant terms in Eq. \eqref{eq:first-integral-outer}. 

Let us call the correction to $v_0$ due to the remaining perturbation terms $v_1$. This function  solves  the equation
\[
2\sin\left(\Theta\right)\frac{\partial v_1}{\partial \Theta} = -2\cos\left(\Theta \right)v_1 - \frac{1}{A^2\cos^2\left(\Theta\right)} - \frac{\partial p}{\partial T}\frac{\log(A)}{A} - 2\frac{1}{A}\frac{\partial p}{\partial T}\log\left[\cos\left(\Theta\right)\right].
\]
When the first term on the right is moved to the left, their sum is a total derivative. Thus the general solution to this equation is readily computed by taking the antiderivative of the remaining terms and dividing by 
$\sin(\Theta)$. The result is
\[
 v_1= a\frac{1}{\sin\left(\Theta\right)} -\frac{1}{2 A^2}\sec\left(\Theta\right)- \frac{1}{2}\frac{\partial p}{\partial T} \frac{\log(A)}{A}\frac{\Theta}{\sin\left(\Theta\right)}  - \frac{1}{\sin(\Theta)}\frac{1}{A}\frac{\partial p}{\partial T}\int_0^{\Theta} \log[\cos(t)]\,dt,
\]
where $a$ is a constant of integration. Symmetry requires $a=0$.

Because we will match $v_1$ to the next order solution to the inner problem, we expand $v_1$ for small 
$\epsilon_{\pm} =\Theta \mp \frac{\pi}{2}$.
We use the estimate
\[
\left(\int_0^{\left(\pm\frac{\pi}{2} + \epsilon_\pm\right)}-\int_0^{\pm\frac{\pi}{2}} \right)\log\left[\cos(t)\right]\,dt\sim \int_0^{\epsilon_\pm} \log\left(\mp t\right)\,dt \sim \epsilon_\pm \log\left(\mp\epsilon_\pm\right) - \epsilon_\pm
\]
 to find
\[
v_1 \sim \pm\frac{1}{2A^2}\frac{1}{\epsilon_\pm} - \frac{\pi}{4}\frac{\log(A)}{ A}\frac{\partial p}{\partial T}  -\frac{1}{A}\frac{\partial p}{\partial T}\int_0^{\frac{\pi}{2}}\log\left[\cos(t)\right]\,dt \pm \left(\frac{1}{12 A^2} + \frac{1}{2}\frac{\partial p}{\partial T}\frac{\log(A)}{A}\right)\epsilon_\pm \mp \frac{1}{A}\frac{\partial p}{\partial T}\epsilon_\pm \log(\epsilon_\pm)+ \cdots
\]
The quadratic and higher order terms in this expansion are negligible provided $|\epsilon_\pm|\ll1$; our principle interest is in the terms which are $O(\epsilon_\pm^0)$. These terms are significant in determining $B_\pm$ in the inner region.

We first expand the corrected leading outer solution as
\[
v_0\sim\cos\left[\sqrt{\beta}\left(\epsilon_\pm \pm \frac{\pi}{2}\right)\right]\sim \mp\left(\epsilon_\pm \pm \frac{\pi}{4}\frac{1}{A}\frac{\partial A}{\partial T}\right) \pm \frac{\epsilon_\pm^3}{6},
\]
and we find that our modification of $v_0$ will not affect the leading order match between $v_0$ and $w_0$ provided $\frac{1}{A}\frac{\partial A}{\partial T}\to 0$. 
Then the full expression for $u$ in the outer region now is 
\begin{multline}
u\sim  \pm \frac{1}{2A^{\frac{3}{2}}}\frac{1}{\epsilon_\pm} - \frac{\pi}{4}\frac{1}{\sqrt{A}}\frac{\partial A}{\partial T}   -  \frac{\pi}{4}\frac{\log(A)}{\sqrt{A}}\frac{\partial p}{\partial T}  -\frac{1}{\sqrt{ A}}\frac{\partial p}{\partial T}\int_0^{\frac{\pi}{2}}\log\left[\cos(s)\right]\,ds \\
 \mp \left(\sqrt{A}  - \frac{1}{2}\frac{\partial p}{\partial T}\frac{\log(A)}{\sqrt{ A}}- \frac{1}{12 A^{\frac{3}{2}}}\right)\epsilon_\pm  \mp \frac{1}{\sqrt{ A}}\frac{\partial p}{\partial T}\epsilon_\pm \log(\epsilon_\pm) \pm \sqrt{A}\frac{\epsilon_\pm^3}{6} + \cdots\label{eq:outer-series}
\end{multline}
which expansion is valid for $|\epsilon_{\pm}|\ll1$. The perturbation series itself is valid if  $|\epsilon_\pm|\gg A^{-1}$.

\subsubsection{Next-to-leading order in the inner region}

As in the outer region, we compute the  leading correction to $w_0$ using perturbation theory. Again, one set of perturbation terms is already apparent in Eq. \eqref{eq:first-integral-inner}; another term arrises from the incompressibility constraint, Eq.  \eqref{eq:xformed-1a}. To understand this latter term, recall that  we arrived at the equation for $g$  using the incompressibility constraint, Eq. \eqref{eq:xformed-2}, to replace  factors of $\frac{\partial \phi}{\partial \Theta}$ with $e^{-2g}$.  To next order, the incompressibility constraint gives
\begin{equation}
\frac{\partial \phi_1}{\partial \Theta} = \left(e^{-2g_1} - 1\right)\frac{\partial \phi_0}{\partial \Theta} + \frac{\partial \phi_0}{\partial T} \frac{\partial g_0}{\partial \Theta} - \frac{\partial g_0}{\partial T}\frac{\partial \phi_0}{\partial \Theta}.\label{eq:phi1-eq}
\end{equation}
where $g_1$ is the first correction to $g_0$ in the inner region. We evaluate the final two terms in the above expression using the leading order expressions
%
%
\begin{equation}
\frac{\partial \phi_0}{\partial \Theta} = \frac{1}{A^{-1}+A(\Theta-B_\pm)^2},
\label{eq:dphi0_dTheta}
\end{equation}
%
\begin{equation}
\frac{\partial g_0}{\partial T} = \frac{1}{2A}\frac{\partial A}{\partial T}  - \left[A(\Theta-B_\pm)\frac{\partial B_\pm}{\partial T}+\frac{1}{A^2}\frac{\partial A}{\partial T} \right]\frac{\partial \phi_0}{\partial \Theta},\label{eq:dg_dT}
\end{equation}
\begin{equation}
\frac{\partial g_0}{\partial \Theta} = A(\Theta-B_\pm)\frac{\partial \phi_0}{\partial \Theta}, 
\end{equation}
and
\begin{equation}
\frac{\partial \phi_0}{\partial T} =\left[\left(\Theta-B_\pm\right)\frac{1}{A}\frac{\partial A}{\partial T}  - \frac{\partial B_\pm}{\partial T}\right] \frac{\partial \phi_0}{\partial \Theta}.\label{eq:dphi0_dT}
\end{equation}
Substituting Eqs. \eqref{eq:dphi0_dTheta} through \eqref{eq:dphi0_dT} into Eq. \eqref{eq:phi1-eq}, we arrive at
\begin{equation}
\frac{\partial \phi_1}{\partial \Theta} =  \left[\frac{1}{2A}\frac{\partial A}{\partial T}+ \left(e^{-2g_1} - 1\right)\right]\frac{\partial \phi_0}{\partial \Theta}.\label{eq:dphi1}
\end{equation}
We can modifying Eq. \eqref{eq:g-eq} to include for the first term in braces in Eq. \eqref{eq:dphi1} by making the replacement 
\[
e^{-2g}\to \left(1+\frac{1}{2A}\frac{\partial A}{\partial T}\right)e^{-2g}.
\]
Keeping the leading terms as $T\to -\infty$, we find a revised  equation for $g$,
\[
-\frac{\partial^2 g}{\partial \Theta^2}  + \left(1 + \frac{1}{A}\frac{\partial A}{\partial T}\right)e^{-4g} - 1 - \left(\frac{\partial g}{\partial \Theta}\right)^2 - \frac{\partial p}{\partial T}\left(1 + \frac{1}{2A}\frac{\partial A}{\partial T}\right)e^{-2g}=0.
\] 
Transforming this equation back to the variables $v$ and $\xi$, we arrive at
\[
\left(\frac{\partial w}{\partial \xi}\right)^2 =1 -\beta w^{-2} - 2\frac{1}{A}\frac{\partial p}{\partial T}\log\left(\frac{w}{\sqrt{A}}\right) - \frac{w^2}{A^2},
\]
where again we have kept only the leading correction, which manifests as the coefficient $\beta$. We can incorporate $\beta$ into our solution by modifying $w_0$ as
\[
w_0 = \sqrt{1+(\xi + b)^2} \to \sqrt{\beta}\sqrt{1+\left(\frac{\xi+b}{\sqrt{\beta}}\right)^2} =  \sqrt{\beta+\left(\xi+b\right)^2}.
\]
For large values of $\xi$
\[
w_0 \sim |\xi+b| + \frac{\beta}{2|\xi+b|} + \cdots,
\]
so our modification of $w_0$ will not affect our leading order match between $w_0$ and $v_0$.

The remaining perturbation terms are treated by adding a term to our modified leading order solution. This term, the function $w_1$, is obtained by solving the problem
\[
2\frac{\xi+ b}{\sqrt{1+ (\xi + b)^2}}\frac{\partial w_1}{\partial \xi} = \frac{2w_1}{[1+(\xi+b)^2]^{\frac{3}{2}}} + \frac{\log(A)}{A}\frac{\partial p}{\partial T} - \frac{1}{A}\frac{\partial p}{\partial T}\log[1+(\xi + b)^2] - \frac{1 + (\xi + b)^2}{A^2}
\]
(which is derived from Eq. \eqref{eq:first-integral-inner} by linearization).
We rewrite this problem as
\[
\frac{\partial w_1}{\partial \xi} = \frac{w_1}{[1+(\xi+b)^2](\xi+b)} + \frac{1}{2}\frac{\sqrt{1+ (\xi + b)^2}}{\xi+b}\left[\frac{\log(A)}{A}\frac{\partial p}{\partial T} - \frac{1}{A}\frac{\partial p}{\partial T}\log[1+(\xi + b)^2] - \frac{1 + (\xi + b)^2}{A^2}\right],
\]
and solve for $w_1$ using the integrating factor
\[
\exp\left(-\int_{r}^{\xi+b} \frac{1}{s(1+ s^2)}\,ds\right) = c(r)\frac{\sqrt{1+(\xi+b)^2}}{\xi+b}.
\]
The lower limit of integration, $r$, is an arbitrary---a different choice simply changes coefficient of proportionality $c(r)$. 
We find
 the general solution 
\begin{multline}
w_1 = \frac{s}{\sqrt{1+s^2}}\left[a -\frac{1}{2}\left(\frac{\log(A)}{A}\frac{\partial p}{\partial T}-\frac{\beta}{A^2}\right)\frac{1}{s} +\frac{1}{2A}\frac{\partial p}{\partial T}\frac{\log(1+s^2)}{s}- 2\frac{1}{A}\frac{\partial p}{\partial T} \arctan(s)\right. \\
\left. + \left(\frac{1}{2}\frac{\log(A)}{A}\frac{\partial p}{\partial T} + \frac{1}{A}\frac{\partial p}{\partial T}  - \frac{\beta}{A^2}\right)s - \frac{1}{2A}\frac{\partial p}{\partial T}s\log(1+s^2) - \frac{s^3}{6A^2}\right], 
\end{multline}
where $s=\xi + b$ and $a$ is an unknown constant.

The asymptotic representation for $w_1$ for large $|s|$ is 
\begin{equation}
w_1\sim -\frac{|s|^3}{6A^2}- \frac{1}{A}\frac{\partial p}{\partial T}|s|\log\left(|s|\right)  + \left(\frac{1}{2}\frac{\log(A)}{A}\frac{\partial p}{\partial T} + \frac{1}{A}\frac{\partial p}{\partial T}  - \frac{1}{A^2}+ \frac{1}{12A^2}\right)|s| \pm a - \frac{1}{A}\frac{\partial p}{\partial T}\pi+\cdots,\quad s\gg1.\label{eq:w_1-asymptote}
\end{equation}
We combine Eq. \eqref{eq:w_1-asymptote} with $w_0$ to obtain an asymptotic representation of the inner approximation for $u$ for large values of $\xi$. We define the variable $\epsilon_\pm = \Theta \pm \frac{\pi}{2}$ to express $u$ as
\begin{multline}
u\sim -\frac{\sqrt{A}}{6}|\epsilon_\pm- B_\pm|^3 - \frac{1}{\sqrt{A}}\frac{\partial p}{\partial T}|\epsilon_\pm-B_\pm|\log\left(|\epsilon_\pm - B_\pm|\right)\\
 + \left(\sqrt{A}-\frac{1}{2}\frac{\log(A)}{\sqrt{A}}\frac{\partial p}{\partial T} + \frac{1}{\sqrt{A}}\frac{\partial p}{\partial T} - \frac{11}{12A^\frac{3}{2}}\right)|\epsilon_\pm-B_\pm|  \pm \frac{a}{\sqrt{A}} -\frac{1}{A^\frac{3}{2}}\frac{\partial p}{\partial T}\pi + \cdots.\label{eq:inner-series}
\end{multline}
This series representation of $u$ is valid if $s\gg1$, i.e. if $|\epsilon_\pm|\gg A^{-1}$, while  perturbation series itself is valid if $|\epsilon_\pm|\ll1$. Hence the inner and outer approximations have an overlap region and we can perform asymptotic matching to determine $B_\pm$.

\subsubsection{Next-to-leading order matching}
We match the inner and outer expressions using the approximations
\[
\beta \sim 1 + \frac{1}{A}\frac{\partial A}{\partial T}, \quad \delta_\pm = 
\sqrt{\beta}\left(\pm\frac{\pi}{2} + \epsilon_\pm\right) \mp\frac{\pi}{2} \sim \pm\frac{\pi}{2}\frac{1}{A}\frac{\partial A}{\partial T} + \epsilon_\pm + \cdots. 
\]
Matching the inner solution given by Eq. \eqref{eq:inner-series} and the outer solution given by Eq. \eqref{eq:outer-series} to leading order as $A\to \infty$ requires 
\[
\pm B_\pm \sim -\frac{\pi}{2}\frac{1}{A}\frac{\partial A}{\partial T} - \frac{\pi}{4}\frac{\log(A)}{A}\frac{\partial p}{\partial T}.
\]
We also find that
\begin{equation}
u\sim \pm\frac{\sqrt{A}}{6}\epsilon_\pm^3 + \left(\sqrt{A} - \frac{1}{2}\frac{\partial p}{\partial T} \frac{\log(A)}{\sqrt{A}}\right)|\epsilon_\pm| \mp \sqrt{A}B_\pm,\quad |B_\pm|\ll|\epsilon_\pm|\ll 1\label{eq:overlap}
\end{equation}
in the matching region, while the leading error is proportional to 
\[
\frac{1}{A^{\frac{3}{2}}}\frac{1}{|\epsilon_\pm|} + \frac{1}{\sqrt{A}}\frac{\partial p}{\partial T}|\epsilon_\pm| \log(|\epsilon_\pm|).
\]
This error is negligible compared to the right side of Eq. \eqref{eq:overlap}  if
\[
\sqrt{A}|\epsilon_\pm|^3 \gg \frac{1}{A^{\frac{3}{2}}}\frac{1}{|\epsilon_\pm|} + \frac{1}{\sqrt{A}}\frac{\partial p}{\partial T}|\epsilon_\pm| \log(|\epsilon_\pm|),
\]
and
\[
\sqrt{A}|B_\pm| \gg \frac{1}{A^{\frac{3}{2}}}\frac{1}{|\epsilon_\pm|} + \frac{1}{\sqrt{A}}\frac{\partial p}{\partial T}|\epsilon_\pm| \log(|\epsilon_\pm|),
\]
Below we show that  $A\sim T^2$, $\frac{\partial p}{\partial T}\sim \frac{T}{\log(-T)}$, and $|B_\pm|\sim \frac{1}{T}$, so both of these inequalities is satisfied if
\[
|\epsilon_\pm|\gg  \frac{1}{\sqrt{T\log(T)}}.
\]

\subsubsection{Satisfying the boundary conditions at the interface}

The final step in constructing our asymptotic solution of Eqs. \eqref{eq:xformed} for large negative $T$ is to satisfy the boundary conditions at the interface given by Eqs. \eqref{eq:interface-bc-mt} in the main text. To perform this calculation we first need expressions for $g$ and $\phi$  to first order in the perturbation parameter $(-T)^{-1}$. Our solution for $g$ is
\begin{subequations} 
\begin{equation}
g\sim \log\left[ \cos\left(\sqrt{\beta}\Theta\right) - \frac{\log(A)}{2A}\frac{\partial p}{\partial T}\frac{\Theta}{\sin \left(\sqrt{\beta}\Theta\right)} \right]+ \frac{1}{2}\log(A),\quad |\epsilon_\pm|\gg \frac{1}{\sqrt{|T|}}\label{eq:g-outer}
\end{equation}
and
\begin{equation}
g\sim\frac{1}{2}\log\left[\frac{\beta}{A}+ A\left(\epsilon_\pm - B_\pm\right)\right] + \log\left[1 + \frac{\log(A)}{2A^2}\frac{\partial p}{\partial T}\frac{A^2 (\epsilon_\pm - B_\pm)^2-1}{A^{-1}+A(\epsilon_\pm - B_\pm)^2} - \frac{A}{6}\frac{|\epsilon_\pm - B_\pm|^4}{A^{-1} + A(\epsilon_\pm - B_\pm)^2}\right],\quad |\epsilon_{\pm}|\ll1.\label{eq:g-inner}
\end{equation}
\label{eq:g-inner-outer}
\end{subequations}
Using these and Eq. \eqref{eq:phi1-eq} we compute $\phi_1$ to first order in perturbation theory as well.  Using the expression for $g_1$, we compute
\[
e^{-2g_1} \sim 1 - \frac{\log(A)}{A^2}\frac{\partial p}{\partial T}\frac{A^2(\epsilon_\pm - B_\pm)^2 - 1}{A^{-1} + A(\epsilon_\pm - B_\pm)^2}
\]
in the inner region.
After substituting this expression into Eq. \eqref{eq:dphi1} we integrate to find  expressions for $\phi$ which is accurate to first order in $(-T)^{-1}$, 
\begin{subequations}
\begin{equation}
\phi \sim \frac{1}{A}\tan(\Theta),\quad |\epsilon_\pm| \gg \frac{1}{\sqrt{|T|}}\label{eq:phi-outer}
\end{equation}
\begin{equation}
\phi \sim \frac{1}{\sqrt{\beta}}\arctan\left[\frac{A}{\sqrt{\beta}}\left(\Theta-B_\pm\right)\right],\quad |\epsilon_\pm|\ll 1.\label{eq:phi-inner}
\end{equation}
\label{eq:phi-inner-outer}
\end{subequations}

Since the two half-spaces comprising the  the elastomer interface only differ in their value of their shear moduli (for example, there is no relative prestress) and by a shift of $\pi$ in the $\Theta$ coordinate, we can transform our solution for one half-space into a solution for the other half space simply by replacing the pressure
\[
p\to \frac{p}{\mu}.
\]
The interface between the half-spaces resides in the inner region of each solution. Because $\Theta$ is periodic, we must enforce continuity of displacement and the stress boundary conditions by taking the $+$ solution for one side of the interface and the $-$ solution for the other side. Letting $A_i$, $B_{i\pm}$, and $\beta_i$ where $i=1,2$ stand for the respective functions $A$, $B_\pm$, and $\beta$  in each half-plane, continuity requires
\begin{equation}
\frac{\beta_1}{A_1} - A_1B_{1+}^2 \sim \frac{\beta_2}{A_2} - A_2B_{2-}^2.\label{eq:continuity-eq}
\end{equation}
The shear stress boundary condition [Eq. \eqref{eq:shear-bc-0}],
\[
\frac{\partial g_1}{\partial \Theta} \sim \mu\frac{\partial g_2}{\partial \Theta}, 
\]
requires 
\begin{equation}
A_1B_{1+} \sim \mu A_2B_{2-}.\label{eq:shear-bc}
\end{equation}
Since we require the periodicity $\phi(\Theta + 2\pi) = \phi(\Theta) + 2\pi$, we find that 
\begin{equation}
\int_0^\pi\frac{\partial \phi}{\partial \Theta}\,d\Theta\sim \frac{\pi}{2\sqrt{\beta_1}} + \frac{1}{\sqrt{\beta_1}}\arctan\left(-\frac{A_1B_{1+}}{\sqrt{\beta_1}}\right) - \frac{1}{\sqrt{\beta_2}}\arctan\left(-\frac{A_2B_{2-}}{\sqrt{\beta_2}}\right) + \frac{\pi}{2\sqrt{\beta_2}} \sim \pi. \label{eq:phi-periodicity}
\end{equation}
Then in view of Eq. \eqref{eq:shear-bc}, we infer that  an asymptotic solution to Eq. \eqref{eq:phi-periodicity} as $T\to -\infty$ can be obtained only if $A_iB_{i\pm}\to \infty$. Expanding Eq. \eqref{eq:phi-periodicity} in light of this fact and presuming $B_{1+}>0$ gives the equation
\begin{equation}
\frac{1}{A_1B_{1+}} -\frac{1}{A_2B_{2-}}+ \frac{\pi}{4}\left(\frac{1}{A_1}\frac{\partial A_1}{\partial T} + \frac{1}{A_2}\frac{\partial A_2}{\partial T}\right)\sim 0.\label{eq:phi-periodicity-2}
\end{equation}
While from Eq. \eqref{eq:continuity-eq} we deduce that
\[
A_1B_{1+}^2\sim A_2B_{2-}^2,
\] 
as well as the relations
\begin{equation}
B_{1+}\sim \frac{B_{2-}}{\mu},\quad A_1\sim \mu^2 A_2.\label{eq:proportionality}
\end{equation}

Because $A_1\propto A_2$, we can use Eqs. \eqref{eq:phi-periodicity-2} and \eqref{eq:shear-bc} to express $B_{1+}$ in terms of $A_1$ and similarly for $B_{2-}$ and $A_2$. We find
\begin{subequations}
\begin{align}
B_{1+} &= \left(1-\mu\right)\left(\frac{\pi}{2}\frac{\partial A_1}{\partial T}\right)^{-1}\label{eq:B1-eq-bc}\\
B_{2-} &= \left(\frac{1}{\mu}-1\right)\left(\frac{\pi}{2}\frac{\partial A_2}{\partial T}\right)^{-1},
\end{align}
\end{subequations}
which are both positive if $\mu<1$, i.e. half-space one is stiffer than half-space two.

We close our system of equations for $A_i$ and $B_{i\pm}$ with the formulae found by asymptotic matching,
\begin{subequations}
\begin{align}
 B_{1+} \sim -\frac{\pi}{2}\frac{1}{A_1}\frac{\partial A_1}{\partial T} - \frac{\pi}{4}\frac{\log(A_1)}{A_1}\frac{\partial p}{\partial T},\label{eq:B1-eq-matching}\\
 B_{2-} \sim \frac{\pi}{2}\frac{1}{A_2}\frac{\partial A_2}{\partial T} + \frac{\pi}{4}\frac{\log(A_2)}{A_2}\frac{1}{\mu}\frac{\partial p}{\partial T}.\label{eq:B2-eq-matching}
\end{align}
\end{subequations}
We use the relations in Eqs. \eqref{eq:proportionality} to rewrite the second equation as
\begin{equation}
 B_{1+} \sim \frac{1}{\mu}\frac{\pi}{2}\frac{1}{A_1}\frac{\partial A_1}{\partial T} + \frac{\pi}{4}\frac{\log\left(\frac{A_1}{\mu^2}\right)}{A_1}\frac{\partial p}{\partial T}.\label{eq:B2-eq-matching-2}
\end{equation}
As we require this equation to hold as $A_1\to \infty$, adding Eqs. \eqref{eq:B1-eq-matching} and \eqref{eq:B2-eq-matching-2} yields
\[
B_{1+}\sim  \frac{\pi}{2}\left(\frac{1}{\mu}-1\right)\frac{1}{2A_1}\frac{\partial A_1}{\partial T}.
\]
Combining this expression with Eq. \eqref{eq:B1-eq-bc} yields a single equation for $A_1$,
\[
\frac{\partial A}{\partial T}\sim\sqrt{\mu \frac{8}{\pi^2}A_1}.
\]
The solution of this equation is
\begin{subequations}
\begin{equation}
A_1 \sim \frac{2\mu}{\pi^2}T^2.\label{eq:A1-formula}
\end{equation}
We immediately compute
\begin{equation}
B_{1+} \sim \frac{\pi}{2}\left(\frac{1}{\mu}-1\right)\frac{1}{T},\quad B_{2-}\sim \frac{\pi}{2}(1-\mu)\frac{1}{T},\quad A_2\sim \frac{2}{\mu\pi^2}T^2\label{eq:AB-formula}
\end{equation}
\label{eq:AB-formula-all}
\end{subequations}
and the equation
\[
 \frac{\partial p}{\partial T} \sim 4\frac{1+\mu}{\pi^2}\frac{-T}{\log(-T)}
\]
for the isotropic part of the pressure.
Integrating this expression for $p$ once results in 
\begin{equation}
p\sim-2\frac{1+\mu}{\pi^2}\frac{T^2}{\log(-T)},\label{eq:p-formula}
\end{equation}
which hold for asymptotically large $-T$.

Finally, we complete the asymptotic solution of Eqs. \eqref{eq:xformed}  at an interface by imposing the normal stress boundary condition given by Eq. \eqref{eq:normal-force-bc}. At the interface,
\[
\frac{\partial \phi}{\partial \Theta} \sim \frac{1}{A_1B_{1+}^2} 
= \frac{2\mu}{(1-\mu)^2}.
\]
Therefore we can satisfy the normal stress boundary condition if the pressure jumps by an amount
\begin{equation}
p_1 - p_2 \sim \frac{2\mu}{1-\mu},\label{eq:pressure-jump}
\end{equation}
that is, the pressure in the stiffer half-space (half-space one) is higher. When we satisfy Eq. \eqref{eq:p-eq}, we find the final expression for the pressure in the inner region of half-space one,
\begin{subequations}
\begin{equation}
p_1 \sim -2\frac{1+\mu}{\pi^2}\frac{T^2}{\log(-T)} - 2\frac{1+\mu}{\pi^2}\frac{T}{\log(-T)}\log\left[\frac{1}{A_1}+A_1\left(\Theta - B_{1\pm}\right)^2\right].\label{eq:pressure-inner}
\end{equation}
and
\begin{equation}
p_1 \sim -2\frac{1+\mu}{\pi^2}\frac{T^2}{\log(-T)} - 4\frac{1+\mu}{\pi^2}T\left[1 +\frac{\log\cos(\Theta)}{\log(-T)}\right]\label{eq:pressure-outer}
\end{equation}
\label{eq:pressure-inner-outer}
\end{subequations}
in the outer region. The pressure in the softer half space differs by a constant as per Eq. (45)

Eqs. \eqref{eq:g-inner-outer} for $g$, Eqs. \eqref{eq:phi-inner-outer}  for $\phi$, and Eqs. (45) 
and 
Eq. (46) 
for $p$, together with the relations in Eqs. \eqref{eq:AB-formula-all}  constitute an asymptotic solution of Eqs. \eqref{eq:xformed} as $T\to -\infty$. Notably, all corrections to these formulae tend to zero as $T \to -\infty$, and $\frac{\partial^2 s}{\partial T^2}$ and $\frac{\partial^2 \phi}{\partial T^2}$ both tend to zero for each $\Theta$ as $T\to -\infty$, which justifies calling this solution scale invariant.

%
Note that while the boundary conditions in Eqs. \eqref{eq:interface-bc-mt} limit to the free surface boundary conditions as $\mu\to 0$ and to a generic interior point as $\mu\to 1$, the pinch solution does not smoothly evolve into a crease or an affine deformation respectively (as can be seen by examine the different rates of divergence of the pressure as $T\to -\infty$ for each solution). These other solutions  likely emerge as intermediate asymptotes as $\mu\to0$ or $\mu\to 1$ and $T\to -\infty$.

\end{document}